# Bulk-plasmon-mediated free-electron radiation beyond the conventional formation time

Fuyang Tay[2,3,#], Xiao Lin[1,4,#,†], Xihang Shi[5,#], Hongsheng Chen[1,4,6], Ido Kaminer[5,†], and Baile Zhang[7,8,†]

[1]Interdisciplinary Center for Quantum Information, State Key Laboratory of Modern Optical Instrumentation, ZJU-Hangzhou Global Science and Technology Innovation Center, College of Information Science and Electronic Engineering, Zhejiang University, Hangzhou 310027, China

[2]Department of Electrical and Computer Engineering, Rice University, Houston, TX 77005, USA

[3]Applied Physics Graduate Program, Smalley–Curl Institute, Rice University, Houston, TX 77005, USA

[4]International Joint Innovation Center, the Electromagnetics Academy at Zhejiang University, Zhejiang University, Haining 314400, China

[5]Department of Electrical Engineering, Technion-Israel Institute of Technology, Haifa 32000, Israel

[6]Jinhua Institute of Zhejiang University, Zhejiang University, Jinhua 321099, China

[7]Division of Physics and Applied Physics, School of Physical and Mathematical Sciences, Nanyang Technological University, 637371, Singapore

[8]Centre for Disruptive Photonic Technologies, Nanyang Technological University, 637371, Singapore

[#]These authors contributed equally to this work.

[†]Corresponding authors. E-mail: xiaolinzju@zju.edu.cn; kaminer@technion.ac.il; blzhang@ntu.edu.sg

**Free-electron radiation is a fundamental photon emission process that is induced by fast-moving electrons interacting with optical media. Historically, it has been understood that, just like any other photon emission process, free-electron radiation must be constrained within a finite time interval known as the "formation time", whose concept is applicable to both Cherenkov radiation and transition radiation, the two basic mechanisms describing radiation from a bulk medium and from an interface, respectively. Here we reveal an alternative mechanism of free-electron radiation far beyond the previously defined formation time. It occurs when a fast electron crosses the interface between vacuum and a plasmonic medium supporting bulk plasmons. While emitted continuously from the crossing point on the interface — thus consistent with the features of transition radiation — the extra radiation beyond the conventional formation time is supported by a long tail of bulk plasmons following the electron's trajectory deep into the plasmonic medium. Such a plasmonic tail mixes surface**

**and bulk effects, and provides a sustained channel for electron-interface interaction. These results also settle the historical debate in Ferrell radiation, regarding whether it is a surface or bulk effect, from transition radiation or plasmonic oscillation.**

The interaction between fast-moving electrons and optical media causes free-electron radiation. A famous example is Cherenkov radiation [1–6], in which photons are emitted from the bulk of a medium when the electron's speed exceeds the speed of light in the medium. Another type of free-electron radiation, known as transition radiation [7–14], refers to photon emission from an interface—when an electron crosses an interface between different media, the electron will always emit photons, at any speed. As an alternative photon emission mechanism apart from atomic spontaneous emission and stimulated emission, free-electron radiation plays a significant role in many practical applications, ranging from high-energy particle detectors, free-electron lasers, electron microscopies, medical imaging, security scanning, to astronomy and cosmology [15–21].

In any type of radiation, it takes time for photons to be emitted, and the time interval is called the formation time. In the context of free-electron radiation, the concept of formation time in a bulk medium was firstly proposed by Ter-Mikaelian in Landau's seminar in 1952 [22,23], and then further developed by Landau himself [24,25], with its physical effects experimentally demonstrated in 1990s [26–28]. Later, Ginzburg extended the formation time concept into photon emission from an interface, namely transition radiation [8,9,29]. As defined by Ginzburg, photons in transition radiation are emitted within one formation time [9]. This concept has already provided valuable guidance for practical applications. For example, the influence of formation time should be avoided in the design of transition radiation detectors [10], which are widely used in the identification of high-energy particles.

Here we reveal an alternative mechanism of free-electron radiation far beyond the previously defined formation time. It can occur when a fast-moving electron crosses the interface between free space and a plasmonic medium supporting bulk plasmons, such as metals at the plasma frequency (Fig. 1). This radiation is supported by a long tail of bulk plasmons following the electron's trajectory deep into the plasmonic medium (see Supplementary Movie S1-S4 for a demonstration). Such a plasmonic tail mixes the bulk and surface effects, making the radiation neither a pure bulk effect as in Cherenkov radiation, nor a pure surface effect as in transition radiation. A striking feature for this bulk-plasmon-mediated radiation, as we will demonstrate later, is its long duration far beyond the formation time historically defined for free-electron radiation.

It is interesting to note that such a geometry of electrons bombarding plasmonic media has been long studied in the field of plasmonics since 1950s. In fact, the existence of surface plasmons was firstly confirmed with electrons bombarding a metal film [30]. However, the complex electron-photon-plasmon interaction has left a few issues in the history that have not been fully resolved. A typical example is the Ferrell radiation [31,32] (i.e., the enhanced radiation at the plasma frequency of the bulk medium), which could be ascribed to a surface effect with pure transition radiation, or a bulk effect induced by plasmonic oscillation, or both, but so far there is no decisive conclusion [33–40] (see supplementary section S5 for a historical survey). With our revealed mechanism of bulk-plasmon-mediated free-electron radiation beyond the conventional formation time, it becomes feasible to settle this historical debate here.

We consider in Fig. 1 that a fast electron moves with a velocity $\bar{v} = \hat{z}v$ and penetrates the interface separating region 1 and region 2, where their relative permittivity is $\varepsilon_{1r} = 1$ (free space) and $\varepsilon_{2r}$ (either a dielectric or a plasmonic medium), respectively, $v = 0.4c$, and $c$ is the speed of light in free space. By imposing the constraint of $\varepsilon_{2r} < (c/v)^2 = 6.25$, we can exclude the possibility of Cherenkov radiation, since the electron's speed falls below the Cherenkov threshold. Consequently, the only possible radiation, according to conventional theories of classical electrodynamics, should be transition radiation, as studied by Ginzburg and many other colleagues [9,12,29,41]. Since we are interested in the backward radiation propagating almost parallel to the electron's trajectory, the conventional formation time of transition radiation, as defined by Ginzburg, is $t_f(\omega) = \frac{2\pi}{\omega\left|1+\frac{v}{c}\sqrt{\varepsilon_{1r}}\right|} + \frac{2\pi}{\omega\left|1-\frac{v}{c}\sqrt{\varepsilon_{2r}}\right|}$ [8,9], where $\omega$ is the angular frequency; see Methods for details. To facilitate the discussion, $t_{f0} = t_f(\omega_p)$ is chosen as a reference to normalize the horizontal coordinate of time in Fig. 2, where $\omega_p$ is the plasma frequency of plasmonic media (e.g. metals). Accordingly, the length of formation zone, also known as the formation length or coherence length [8,22,23,42], is $L_{f0} = vt_{f0}$.

Now we consider a plasmonic medium in region 2 with $\varepsilon_{2r} = \varepsilon_{\text{Drude}}(\omega)$. To capture the role of losses, we employ a Drude-like formula to describe the relative permittivity of plasmonic media, namely $\varepsilon_{\text{Drude}}(\omega) = 1 - \frac{\omega_p^2}{\omega^2 + i\omega/\tau}$, where the plasma frequency is set to be $\omega_p = 13.9$ petahertz, and $\tau$ is the relaxation time; note that the nonlocal response of plasmonic media has a minor influence on the radiation revealed here; see Figs. S7-S8. As a comparison, we also consider the control situation when region 2 is a regular dielectric with $\varepsilon_{2r} = 2$.

Figure 2a-f shows the temporal evolution of the backward-radiation field at a point $\bar{r}_{\text{far}}$ far away from the interface but close to the electron's trajectory, where the angle between $\bar{r}_{\text{far}}$ and $-\bar{z}$ is $\theta_{\text{far}} \simeq 7°$. The major part of emitted photons at $\theta_{\text{far}}$ is formed within $t_{f0}$, as shown in Fig. 2a & d. If the fast electron is within the conventional formation zone of transition radiation, according to Ginzburg's analysis, the electron can directly interact with the interface. Then the significant process of photon emission within the conventional formation time in Fig. 2a & d can be treated as a surface effect. Beyond $t_{f0}$, if region 2 is a plasmonic medium with $\varepsilon_{2r} = \varepsilon_{\text{Drude}}(\omega)$, there are sizable radiation fields that oscillate periodically far beyond $t_{f0}$ in Fig. 2a-b, where the relaxation time $\tau = \tau_0$ is used and $\tau_0 = 700/\omega_p$. In contrast, if region 2 is a dielectric with $\varepsilon_{2r} = 2$, the radiation field beyond $t_{f0}$ is negligible as shown in Fig. 2d-e.

Moreover, the radiation fields beyond the conventional formation time $t_{f0}$ in Fig. 2b have their frequency close to the plasma frequency. Figure 2c shows that after the electron crosses the interface, the radiation peak around the plasma frequency is mainly contributed by the emission process of photons beyond the conventional formation time $t_\text{f}(\omega \approx \omega_\text{p}) \approx t_\text{f0}$.

Figure 3a-d shows that the revealed extra radiation beyond $t_\text{f0}$ is caused by the interaction between the interface and a long tail of excited bulk plasmons (both the transverse and longitudinal electromagnetic waves are considered for bulk plasmons; see supplementary section S7). The reason is that the tail of excited bulk plasmons in the plasmonic medium not only follows the electron's trajectory but also can attach to (and thus interact with) the interface for a long time (Fig. 3a). In other words, even when the fast electron is far beyond the formation zone of transition radiation, the electron will continue to interact with the interface at the same crossing point; this way, the radiation process is unfinished beyond the conventional formation time, and the fast electron will continue to emit photons from the interface (Fig. 3a). Since the excitation of bulk plasmons is a bulk response of the plasmonic medium, it is reasonable to treat the formation of the radiation beyond $t_\text{f0}$ as a mixture of surface-bulk effect, instead of solely a surface effect. Besides, since the surface plasmons at the surface of plasmonic media exist far below the plasma frequency $\omega_\text{p}$ (i.e., $\omega < \omega_\text{p}/\sqrt{2}$) but the excited bulk plasmons have their frequency close to $\omega_\text{p}$, the surface plasmons cannot be excited by the bulk plasmons but can be excited only by the direct electron-interface interaction within the conventional formation time [29]; see more discussion about Fig. 3a-d in Methods. This way, we denote the radiation beyond $t_\text{f0}$ as the bulk-plasmon-mediated free-electron radiation, while denoting the radiation within $t_\text{f0}$ as the conventional free-electron radiation (or transition radiation).

Due to the interface-bulk effect, we shall re-define the formation time for this bulk-plasmon-mediated free-electron radiation. This formation time, labelled as $T_{e-\text{interface}}$, describes the electron-interface interaction time and is here defined as the time taken for the tail of bulk plasmons to leave the interface, which can be treated simply as the ratio between the full length of the tail (created by a fast electron moving inside a homogeneous plasmonic medium) and the electron's velocity. Apparently, this formation time $T_{e-\text{interface}}$ is highly dependent on the loss of plasmonic media. For example, $T_{e-\text{interface}} > 100 t_{\text{f0}}$ if $\tau > 0.1\tau_0$, as shown in Fig. 3e. The large value of $T_{e-\text{interface}}$ directly indicates that the bulk-plasmon-mediated free-electron radiation can occur far beyond the conventional formation time of transition radiation. This bulk-plasmon-mediated free-electron radiation is distinct from the conventional free-electron radiation trapped by resonators (inside which the out-coupling process of the already-formed photons can last for a relatively-long time without the existence of electron-resonator interactions, if the quality factor of resonators is high) [43]; see Methods.

To further understand the bulk-plasmon-mediated radiation, we explore it in the frequency domain in Fig. 4. Figure 4a-b shows the angular spectral energy density $U_1(\omega, \theta)$ of backward radiation. From Fig. 4a, there is a radiation peak near the plasma frequency if region 2 is the plasmonic medium. This radiation peak is historically known as the Ferrell radiation, [11,31–40,44] which has caused a long debate about its physical origin, regarding whether it is a surface or bulk effect, from transition radiation or plasmonic oscillation (see supplementary section S5 for a detailed histrocial survey). Nowadays, most literatures have ascribed Ferrell radiation to radiative surface plasmons or the so-called Ferrell mode [32,38,40,44], but a complete picture is still lacking to fully settle the debate.

We find that the introduction of the bulk-plasmon-mediated free-electron radiation beyond the conventional formation time can address this issue. Firstly, there are two types of plasmonic oscillations simultaneously, bulk plasmons in the bulk and radiative surface plasmons on the interface. While bulk plasmons are excited as the plasmonic tail deep into the plasmonic medium, the tail also touches the interface, exciting radiative surface plasmons. Secondly, while it is the radiative surface plasmons that directly emit photons from the interface (but that alone cannot continuously emit photons far beyond the conventional formation time due to their large decay rate), it is the bulk plasmons that provide the energy, and the bulk plasmons in turn draw energy from the fast-moving electron. Along the way, the electromagnetic field extends coherently from the fast-moving electron to the interface. In other words, it takes a long distance, and thus a long time interval, to "peel off" the electromagnetic field from the electron, eventually forming photons. Thirdly, the Ferrell radiation is contributed by both transition radiation and plasmonic oscillation. These two contributions are impossible to be distinguished in the frequency domain but can be separated in the time domain. The transition radiation, as described by Ginzburg, occurs only within one conventional formation time. The significant radiation beyond the conventional formation time comes from the plasmonic oscillation, as we have analyzed in Fig. 2.

In conclusion, we have revealed the emergence of bulk-plasmon-mediated free-electron radiation beyond the conventional formation time by investigating the penetration of a fast-moving electron through the interface of a plasmonic medium. This extra radiation is closely related to a long tail of bulk plasmons, which follows the electron's trajectory deep into the plasmonic medium and can attach to and thus interact with the interface for a very long time. Correspondingly, this tail provides a unique route to mix the surface and bulk effects, significantly extend the electron-interface interaction, and then create light emission far beyond the conventional formation time. Therefore, the revealed bulk-plasmon-mediated free-electron radiation is intrinsically a mixture of surface-bulk effect, distinct from Cherenkov radiation as purely a bulk effect or transition radiation as purely a surface effect. Our finding also provides a new perspective for Ferrell radiation and settles its historical debate regarding its physical origin. The bulk-plasmon-mediated free-electron radiation may further trigger many open questions, concerning, for example, the observation of the long tail of bulk plasmons or the radiation beyond the conventional formation time, the possibility to largely enhance particle-light-matter interactions and the design of advanced light sources by exploiting bulk plasmons in other types of free-electron radiation such as Smith-Purcell radiation and synchrotron radiation.

**Acknowledgements**
X.L. acknowledges the support partly from the National Natural Science Fund for Excellent Young Scientists Fund Program (Overseas) of China, the National Natural Science Foundation of China (NSFC) under Grant No. 62175212, Zhejiang Provincial Natural Science Fund Key Project under Grant No. LZ23F050003, and the Fundamental Research Funds for the Central Universities (2021FZZX001-19). H.C. acknowledges the support from the Key Research and Development Program of the Ministry of Science and Technology under Grants No. 2022YFA1404704, 2022YFA1404902, and 2022YFA1405200, the National Natural Science Foundation of China (NNSFC) under Grants No.11961141010 and No. 61975176. X.S. is supported in part by a fellowship of the Israel Council for Higher Education and by the Technion's Helen Diller Quantum Center. I.K. acknowledges the supported by the ISF (Grant No. 3334/19). B.Z. acknowledges the support from National Research Foundation Singapore Competitive Research Program no. NRF-CRP23-2019-0007, and Singapore Ministry of Education Academic Research Fund Tier 3 under grant no. MOE2016-T3-1-006 and Tier 2 under grant no. MOE2019-T2-2-085.

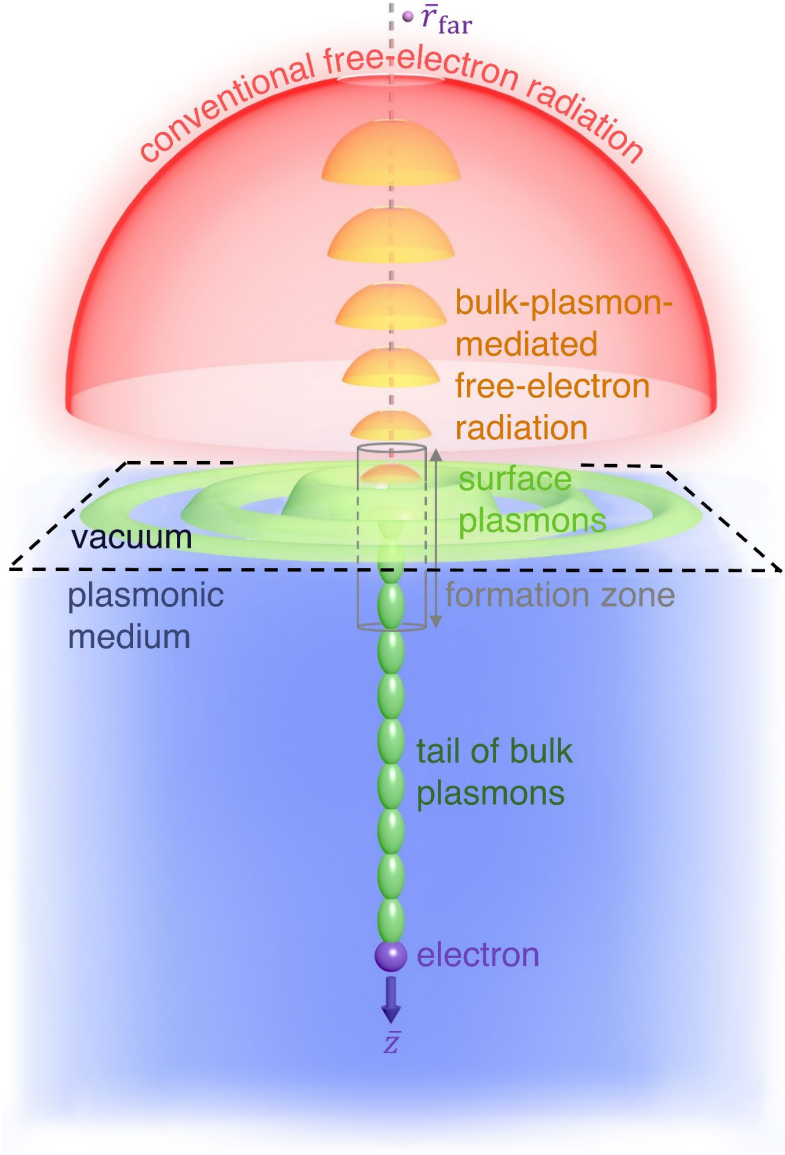

**Fig. 1. Schematic of the bulk-plasmon-mediated free-electron radiation beyond the conventional formation time.** A fast-moving electron with velocity $\bar{v} = \hat{z}v$ impinges on an interface and induces radiation, where $v = 0.4c$ is used. The interface separates regions 1 (namely vacuum) and region 2 (a dielectric or a plasmonic medium), whose relative permittivity is $\varepsilon_{1r}$ and $\varepsilon_{2r}$, respectively. The radiation beyond the conventional formation time occurs due to the formation of a long tail of bulk plasmons when the electron moves inside the plasmonic medium, whose optical response is described by a Drude-like formula, namely $\varepsilon_{\text{Drude}}(\omega) = 1 - \frac{\omega_p^2}{\omega^2 + i\omega/\tau}$.

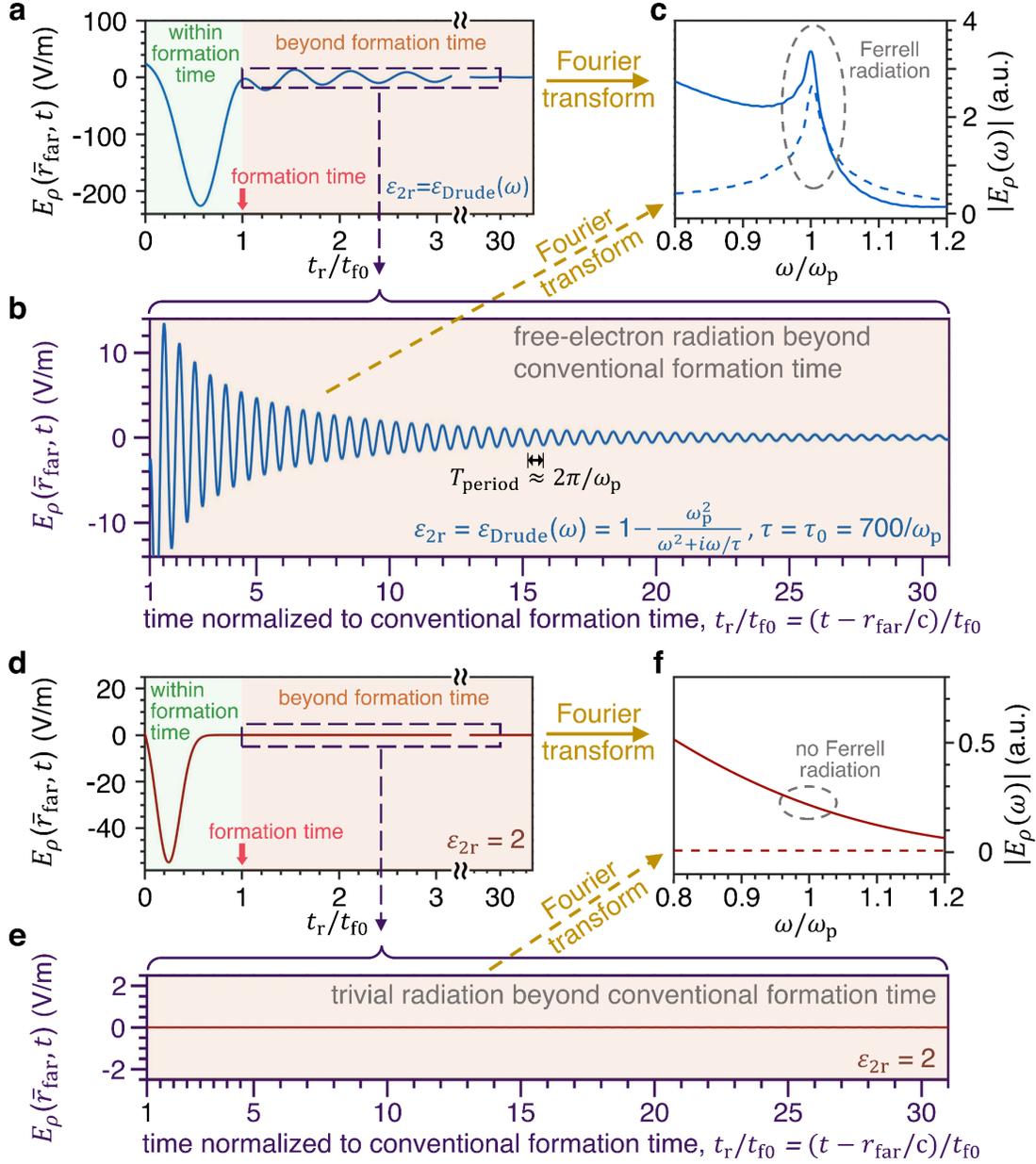

**Fig. 2. Representation in the time domain of the bulk-plasmon-mediated free-electron radiation beyond the conventional formation time.** The structural setup is the same as Fig. 1. (**a**) Temporal evolution of the backward-radiation field $E_\rho(\bar{r}_{\text{far}}, t)$ at a point $\bar{r}_{\text{far}}$ far away from the interface but close to the electron trajectory when region 2 is filled by a plasmonic medium with $\varepsilon_{2r} = \varepsilon_{\text{Drude}}(\omega)$. Here and below, we set $t = 0$ as the moment when the electron enters the formation zone in vacuum, $t_r = t - r_{\text{far}}/c$ as the retarded time, $r_{\text{far}} = |\bar{r}_{\text{far}}|$, $t_{f0} = t_f(\omega = \omega_p)$,

and $t_\text{f}(\omega)$ is the conventional formation time defined for transition radiation. (**b**) An enlarged plot of the radiation field beyond the conventional formation time (highlighted by a purple dashed square in (a)). (**c**) Frequency-domain analysis of the radiation in different ranges of time. The solid line corresponds to the Fourier transform of radiation from $t/t_\text{f0} = 0$ to $t/t_\text{f0} = 40$ in (a), while the dashed line corresponds to the Fourier transform of radiation from $t/t_\text{f0} = 1$ to $t/t_\text{f0} = 40$ in (b). (**d-f**) Conventional free-electron radiation when region 2 is filled by a regular dielectric with $\varepsilon_\text{2r} = 2$. All analysis in (d-f) is the same as (a-c).

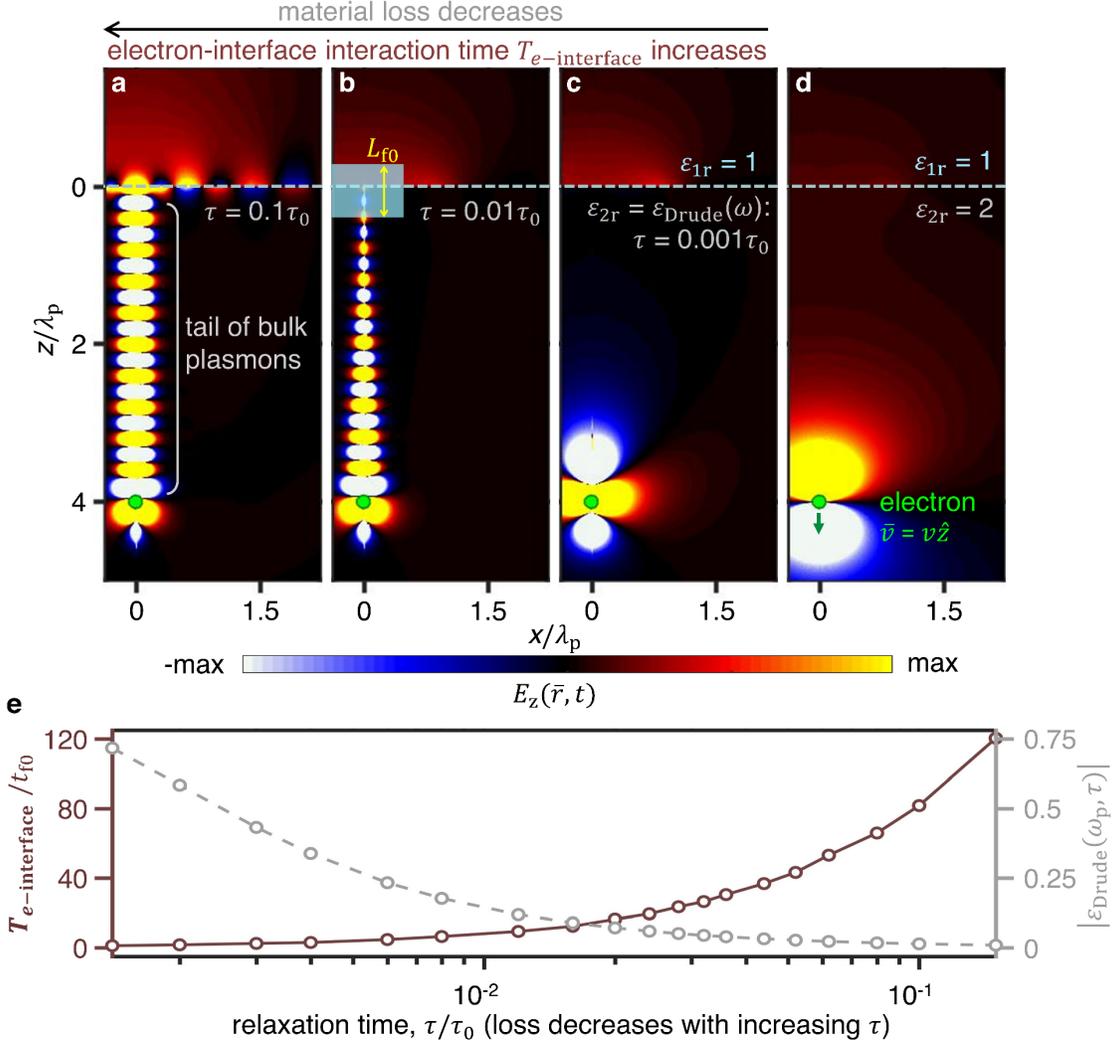

**Fig. 3. Spacetime-domain illustration of the underlying physics for the bulk-plasmon-mediated free-electron radiation.** This radiation is caused by the indirect electron-interface interaction, which is mediated via a long tail of bulk plasmons and can exist far beyond the conventional formation time. (**a-d**) Spatial distribution of the electric field $E_z(\vec{r}, t)$; see also in Movies S1-S4. (**e**) Electron-interface interaction time $T_{e-\text{interface}}$ as a function of the relaxation time $\tau$. $T_{e-\text{interface}}$ can far exceed $t_{f0}$, if $\tau$ increases or if $|\varepsilon_{\text{Drude}}(\omega_p, \tau)|$ (instead of merely $\text{Im}(\varepsilon_{\text{Drude}}(\omega_p, \tau))$) decreases down to a minor value. The shaded region at the interface in (b) represents the formation length of transition radiation, $L_{f0}$. Besides, $\lambda_p = 2\pi c/\omega_p$.

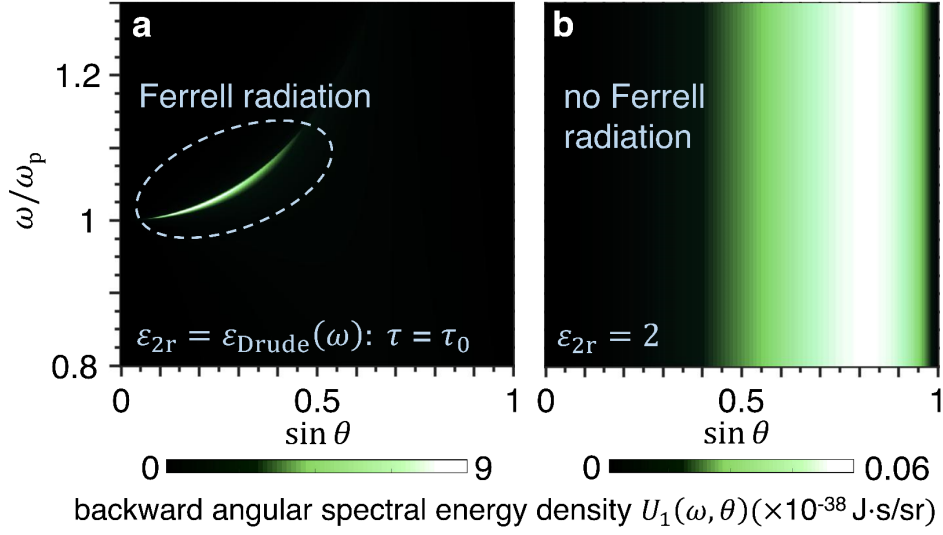

**Fig. 4. Frequency-domain analysis of the bulk-plasmon-mediated free-electron radiation.** The structural setup is the same as Fig. 1. (**a, b**) Angular spectral energy density $U_1(\omega,\theta)$ of the backward radiation. Region 2 is filled by a plasmonic medium with $\varepsilon_{2r} = \varepsilon_{\text{Drude}}(\omega)$ in (a) and a regular dielectric with $\varepsilon_{2r} = 2$ in (b).

Supplementary Material for

# Bulk-plasmon-mediated free-electron radiation beyond the conventional formation time

Fuyang Tay, Xiao Lin, Xihang Shi, Hongsheng Chen, Ido Kaminer, and Baile Zhang

**This PDF file includes:**

-- Supplementary Methods

    Section S1. Derivation of free-electron radiation from an interface

    Section S2. Conventional formation time and formation zone

    Section S3. Angular spectral energy density and energy spectral density of backward free-electron radiation

    Section S4. More discussion on the time-domain study of the bulk-plasmon-mediated free-electron radiation from the vacuum-plasmonic medium interface

    Section S5. History of Ferrell radiation

    Section S6. Influence of the electron's velocity on the long tail of bulk plasmons

    Section S7. Influence of the nonlocal response of plasmonic media and the longitudinal waves on the bulk-plasmon-mediated free-electron radiation

-- Supplementary References

-- Caption for Supplementary Video

# Supplementary Methods

## Section S1. Derivation of free-electron radiation from an interface

By following Ginzburg and Frank's theory of transition radiation [7-9,12], below we analytically derive the radiation process when a swift electron perpendicularly crosses an interface. The basic setup is the same as that shown in Fig. 1 in the main text.

The current density for a swift particle with a charge $q$ and a velocity $\bar{v} = \hat{z}v$ is given as

$$\bar{J}^q(\bar{r}, t) = \hat{z}qv\delta(x)\delta(y)\delta(z - vt). \tag{1}$$

We define the coordinate perpendicular to the boundary as $\bar{r}_\perp = \hat{x}x + \hat{y}y$. The current density is decomposed into a series of the function of frequencies and wave vectors via Fourier transform,

$$\bar{J}^q(\bar{r}, t) = \hat{z} \int j^q_{\bar{\kappa}_\perp,\omega}(z) e^{i(\bar{\kappa}_\perp \cdot \bar{r}_\perp - \omega t)} d\bar{\kappa}_\perp d\omega, \tag{2}$$

where $\bar{\kappa}_\perp = \hat{x}\kappa_x + \hat{y}\kappa_y$. From equations (1-2), one gets $j^q_{\bar{\kappa}_\perp,\omega}(z) = \frac{q}{(2\pi)^3} \exp\left(i\frac{\omega}{v}z\right)$. Moreover, the electromagnetic fields can also be decomposed by Fourier transform,

$$\bar{E}(\bar{r}, t) = \int \bar{E}_{\bar{\kappa}_\perp,\omega}(z) e^{i(\bar{\kappa}_\perp \cdot \bar{r}_\perp - \omega t)} d\bar{\kappa}_\perp d\omega,$$

$$\bar{H}(\bar{r}, t) = \int \bar{H}_{\bar{\kappa}_\perp,\omega}(z) e^{i(\bar{\kappa}_\perp \cdot \bar{r}_\perp - \omega t)} d\bar{\kappa}_\perp d\omega. \tag{3}$$

Since all the quantities are decomposed, the indices $\bar{\kappa}_\perp$ and $\omega$ are omitted for the sake of simplicity. By solving Maxwell equations, one can obtain the expression for the component of the electric field parallel to the particle's trajectory,

$$\frac{\partial^2}{\partial z^2}(\varepsilon_r E_z) + \varepsilon_r \left(\frac{\omega^2}{c^2}\varepsilon_r - \kappa_\perp^2\right) E_z = -\frac{iq\omega\mu_0}{(2\pi)^3}\left(\varepsilon_r - \frac{c^2}{v^2}\right) \exp\left(i\frac{\omega}{v}z\right), \tag{4}$$

where $\varepsilon_r$ is the relative permittivity and $c$ is the speed of light in the free space. Only the TM (p-polarized) wave is excited in the studied structures in this work. The solution for equation (4) is a linear sum of the field induced by the moving charge ($E_z^q$) and the freely radiated field ($E_z^R$). One may obtain the solution as

$$E_z^q = -\frac{iq}{\omega\varepsilon_0(2\pi)^3} \frac{1 - \frac{c^2}{v^2\varepsilon_r}}{\varepsilon_r - \frac{c^2}{v^2} - \frac{\kappa_\perp^2 c^2}{\omega^2}} e^{i\frac{\omega}{v}z}, \quad (5)$$

$$E_z^R = \frac{iq}{\omega\varepsilon_0(2\pi)^3} \cdot a \cdot e^{\pm i\left(\frac{\omega}{c}\sqrt{\varepsilon_r - \frac{\kappa_\perp^2 c^2}{\omega^2}}\right)z}. \quad (6)$$

where $a$ is the amplitude of the radiation fields. As the radiation propagates away from the boundary, " + " sign is used for region 2 ($z > 0$) while " - " sign is used for region 1 ($z < 0$). The value of $a$ can be obtained by matching the boundary conditions below,

$$\hat{n} \times (\bar{H}_{1\perp} - \bar{H}_{2\perp})|_{z=0} = 0, \quad \hat{n} \times (\bar{E}_{1\perp} - \bar{E}_{2\perp})|_{z=0} = 0, \quad (7)$$

After some calculations, the amplitudes of the radiation fields in both regions are given as

$$a_{21}^- = \frac{\frac{v}{c}\frac{\kappa_\perp^2 c^2}{\omega^2 \varepsilon_{1r}}(\varepsilon_{2r} - \varepsilon_{1r})\left[1 - \frac{v^2}{c^2}\varepsilon_{1r} + \frac{v}{c}\sqrt{\varepsilon_{2r} - \frac{\kappa_\perp^2 c^2}{\omega^2}}\right]}{\left(1 - \frac{v^2}{c^2}\varepsilon_{1r} + \frac{\kappa_\perp^2 v^2}{\omega^2}\right)\left(1 + \frac{v}{c}\sqrt{\varepsilon_{2r} - \frac{\kappa_\perp^2 c^2}{\omega^2}}\right)\left[\varepsilon_{1r}\sqrt{\varepsilon_{2r} - \frac{\kappa_\perp^2 c^2}{\omega^2}} + \varepsilon_{2r}\sqrt{\varepsilon_{1r} - \frac{\kappa_\perp^2 c^2}{\omega^2}}\right]}, \quad (8)$$

$$a_{12}^+ = \frac{\frac{v}{c}\frac{\kappa_\perp^2 c^2}{\omega^2 \varepsilon_{2r}}(\varepsilon_{2r} - \varepsilon_{1r})\left[1 - \frac{v^2}{c^2}\varepsilon_{2r} - \frac{v}{c}\sqrt{\varepsilon_{1r} - \frac{\kappa_\perp^2 c^2}{\omega^2}}\right]}{\left(1 - \frac{v^2}{c^2}\varepsilon_{2r} + \frac{\kappa_\perp^2 v^2}{\omega^2}\right)\left(1 - \frac{v}{c}\sqrt{\varepsilon_{1r} - \frac{\kappa_\perp^2 c^2}{\omega^2}}\right)\left[\varepsilon_{1r}\sqrt{\varepsilon_{2r} - \frac{\kappa_\perp^2 c^2}{\omega^2}} + \varepsilon_{2r}\sqrt{\varepsilon_{1r} - \frac{\kappa_\perp^2 c^2}{\omega^2}}\right]}. \quad (9)$$

Note that $a_{21}^-$ represents the backward radiation into region 1 while $a_{12}^+$ represents the forward radiation into region 2. Other components of the electromagnetic fields can all be calculated from $E_z$. One can then get the expression for the electromagnetic field in the real space-time domain by integrating the solutions of equations (5-6) concerning $\omega$ and $\kappa_\perp$. Since our model follows the cylindrical symmetry, the solution can be expressed in cylindrical coordinates $(\rho, \phi, z)$ for simplicity,

$$\bar{E}_1^q(\bar{r},t) = \hat{z}\int_{-\infty}^{+\infty} d\omega \frac{-q}{8\pi\omega\varepsilon_0\varepsilon_{1r}}\left(\frac{\omega^2}{c^2}\varepsilon_{1r} - \frac{\omega^2}{v^2}\right)H_0^{(1)}\left(\rho\sqrt{\frac{\omega^2}{c^2}\varepsilon_{1r} - \frac{\omega^2}{v^2}}\right)e^{i\left(\frac{\omega}{v}z-\omega t\right)}$$

$$+\hat{\rho}\int_{-\infty}^{+\infty} d\omega \frac{-q}{8\pi\omega\varepsilon_0\varepsilon_{1r}}\left(i\frac{\omega}{v}\right)\left(-\sqrt{\frac{\omega^2}{c^2}\varepsilon_{1r} - \frac{\omega^2}{v^2}}\right)H_1^{(1)}\left(\rho\sqrt{\frac{\omega^2}{c^2}\varepsilon_{1r} - \frac{\omega^2}{v^2}}\right)e^{i\left(\frac{\omega}{v}z-\omega t\right)}, \qquad (10)$$

$$\bar{H}_1^q(\bar{r},t) = \hat{\phi}\int_{-\infty}^{+\infty} d\omega \frac{iq}{8\pi}\sqrt{\frac{\omega^2}{c^2}\varepsilon_{1r} - \frac{\omega^2}{v^2}}\,H_1^{(1)}\left(\rho\sqrt{\frac{\omega^2}{c^2}\varepsilon_{1r} - \frac{\omega^2}{v^2}}\right)e^{i\left(\frac{\omega}{v}z-\omega t\right)}, \qquad (11)$$

$$\bar{E}_1^R(\bar{r},t) = \hat{z}\int_{-\infty}^{+\infty} d\omega \int_0^{+\infty} d\kappa_\perp \frac{iq}{(2\pi)^3\omega\varepsilon_0} a_{21}^-\kappa_\perp\left(2\pi J_0(\kappa_\perp\rho)\right)e^{i\left[-\left(\frac{\omega}{c}\sqrt{\varepsilon_{1r} - \frac{\kappa_\perp^2 c^2}{\omega^2}}\right)z-\omega t\right]}$$

$$+\hat{\rho}\int_{-\infty}^{+\infty} d\omega \int_0^{+\infty} d\kappa_\perp \frac{iq}{(2\pi)^3\omega\varepsilon_0} a_{21}^-\left(\frac{\omega}{c}\sqrt{\varepsilon_{1r} - \frac{\kappa_\perp^2 c^2}{\omega^2}}\right)\left(i2\pi J_1(\kappa_\perp\rho)\right)e^{i\left[-\left(\frac{\omega}{c}\sqrt{\varepsilon_{1r} - \frac{\kappa_\perp^2 c^2}{\omega^2}}\right)z-\omega t\right]}, \qquad (12)$$

$$\bar{H}_1^R(\bar{r},t) = \hat{\phi}\int_{-\infty}^{+\infty} d\omega \int_0^{+\infty} d\kappa_\perp \frac{iq}{(2\pi)^3\omega\varepsilon_0} a_{21}^-(-\omega\varepsilon_{1r}\varepsilon_0)\left(i2\pi J_1(\kappa_\perp\rho)\right)e^{i\left[-\left(\frac{\omega}{c}\sqrt{\varepsilon_{1r} - \frac{\kappa_\perp^2 c^2}{\omega^2}}\right)z-\omega t\right]}, \qquad (13)$$

$$\bar{E}_2^q(\bar{r},t) = \hat{z}\int_{-\infty}^{+\infty} d\omega \frac{-q}{8\pi\omega\varepsilon_0\varepsilon_{2r}}\left(\frac{\omega^2}{c^2}\varepsilon_{2r} - \frac{\omega^2}{v^2}\right)H_0^{(1)}\left(\rho\sqrt{\frac{\omega^2}{c^2}\varepsilon_{2r} - \frac{\omega^2}{v^2}}\right)e^{i\left(\frac{\omega}{v}z-\omega t\right)}$$

$$+\hat{\rho}\int_{-\infty}^{+\infty} d\omega \frac{-q}{8\pi\omega\varepsilon_0\varepsilon_{2r}}\left(i\frac{\omega}{v}\right)\left(-\sqrt{\frac{\omega^2}{c^2}\varepsilon_{2r} - \frac{\omega^2}{v^2}}\right)H_1^{(1)}\left(\rho\sqrt{\frac{\omega^2}{c^2}\varepsilon_{2r} - \frac{\omega^2}{v^2}}\right)e^{i\left(\frac{\omega}{v}z-\omega t\right)}, \qquad (14)$$

$$\bar{H}_2^q(\bar{r},t) = \hat{\phi}\int_{-\infty}^{+\infty} d\omega \frac{iq}{8\pi}\sqrt{\frac{\omega^2}{c^2}\varepsilon_{2r} - \frac{\omega^2}{v^2}}\,H_1^{(1)}\left(\rho\sqrt{\frac{\omega^2}{c^2}\varepsilon_{2r} - \frac{\omega^2}{v^2}}\right)e^{i\left(\frac{\omega}{v}z-\omega t\right)}, \qquad (15)$$

$$\bar{E}_2^R(\bar{r},t) = \hat{z}\int_{-\infty}^{+\infty} d\omega \int_0^{+\infty} d\kappa_\perp \frac{iq}{(2\pi)^3\omega\varepsilon_0} a_{12}^+\kappa_\perp\left(2\pi J_0(\kappa_\perp\rho)\right)e^{i\left[+\left(\frac{\omega}{c}\sqrt{\varepsilon_{2r} - \frac{\kappa_\perp^2 c^2}{\omega^2}}\right)z-\omega t\right]}$$

$$+\hat{\rho}\int_{-\infty}^{+\infty} d\omega \int_0^{+\infty} d\kappa_\perp \frac{iq}{(2\pi)^3\omega\varepsilon_0} a_{12}^+\left(-\frac{\omega}{c}\sqrt{\varepsilon_{2r} - \frac{\kappa_\perp^2 c^2}{\omega^2}}\right)\left(i2\pi J_1(\kappa_\perp\rho)\right)e^{i\left[+\left(\frac{\omega}{c}\sqrt{\varepsilon_{2r} - \frac{\kappa_\perp^2 c^2}{\omega^2}}\right)z-\omega t\right]}, \qquad (16)$$

$$\bar{H}_2^R(\bar{r},t) = \hat{\phi}\int_{-\infty}^{+\infty} d\omega \int_0^{+\infty} d\kappa_\perp \frac{iq}{(2\pi)^3\omega\varepsilon_0} a_{12}^+(-\omega\varepsilon_{2r}\varepsilon_0)\left(i2\pi J_1(\kappa_\perp\rho)\right)e^{i\left[+\left(\frac{\omega}{c}\sqrt{\varepsilon_{2r} - \frac{\kappa_\perp^2 c^2}{\omega^2}}\right)z-\omega t\right]}. \qquad (17)$$

## Section S2. Conventional formation time and formation zone

For transition radiation, according to Ginzburg's work [8,9], the length of the formation zone is defined as the distance that the charge field and the radiation field separate from each other. In other words, the contribution of the interference term $E^q \cdot E^R$ to the total field energy (proportional to $|E^q + E^R|^2$) must be very small outside the formation zone. According to equations (5-6), the charge field and radiation field for the backward radiation will have a minor interference if [8,9]

$$\left[\frac{\omega}{v} + \frac{\omega}{c}\sqrt{\varepsilon_{1r} - \frac{\kappa_\perp^2 c^2}{\omega^2}}\right] \cdot z \gg 2\pi. \tag{18}$$

Namely, when the interference term oscillates rapidly, its integration over the real space is small. Therefore, the formation length $L_{f1}$ in region 1 and $L_{f2}$ in region 2 are defined as [8,9]

$$L_{f1,2} = \frac{2\pi}{\left|\frac{\omega}{v} \pm \frac{\omega}{c}\sqrt{\varepsilon_{1r,2r} - \frac{\kappa_\perp^2 c^2}{\omega^2}}\right|}, \tag{19}$$

where the "+" sign is used for region 1 while the "−" sign is used for region 2. Accordingly, the formation length $t_{f1}$ in region 1 and $t_{f2}$ in region 2 is defined as $t_{f1,2} = L_{f1,2}/v$, or

$$t_{f1,2} = \frac{2\pi/v}{\left|\frac{\omega}{v} \pm \frac{\omega}{c}\sqrt{\varepsilon_{1r,2r} - \frac{\kappa_\perp^2 c^2}{\omega^2}}\right|}. \tag{20}$$

In this work, we are interested in the backward-radiation fields emitted at small angles (namely $\kappa_\perp \approx 0$) at the frequency very close to the plasma frequency. This way, we denote in the main text $t_f(\omega) = \frac{2\pi}{\omega\left|1+\frac{v}{c}\sqrt{\varepsilon_{1r}}\right|} + \frac{2\pi}{\omega\left|1-\frac{v}{c}\sqrt{\varepsilon_{2r}}\right|}$ as the conventional formation time of transition radiation (whose related fields propagate parallel to the electron's trajectory). To facilitate related discussions, we adopt the conventional formation time $t_{f0} = t_f(\omega_p) = \frac{2\pi}{\omega_p\left|1+\frac{v}{c}\right|} + \frac{2\pi}{\omega_p\left|1-\frac{v}{c}\sqrt{\varepsilon_{2r}}\right|}$ (or the formation length $L_{f0} = t_{f0}v$) as a reference to normalize related parameter in all related figure plots.

## Section S3. Angular spectral energy density and energy spectral density of backward free-electron radiation

*Analytical derivation of angular spectral energy density and energy spectral density*

One can calculate the total energy of emitted photons in the backward free-electron radiation with a simple approach by considering the radiation fields at $t \to \infty$ [12,29,45]. The backward radiation energy in region 1 (or free space) is expressed as

$$W_1 = \int d\bar{r}_\perp \int_0^{+\infty} \varepsilon_1 |\bar{E}_1^R(\bar{r},t)|^2 \, dz, \tag{21}$$

where $|\bar{E}_1^R(\bar{r},t)|^2 = \int\int\int\int \bar{E}_{1|\bar{\kappa}_\perp,\omega}^R(z) \cdot \bar{E}_{1|\bar{\kappa}_\perp',\omega'}^{R*}(z) \exp(i[(\bar{\kappa}_\perp - \bar{\kappa}_\perp') \cdot \bar{r}_\perp - (\omega - \omega')t]) \, d\bar{\kappa}_\perp d\bar{\kappa}_\perp' d\omega d\omega'$, it is the modulus square of the radiation field. Two expressions are used for the integration over the space,

$$\int d\bar{r}_\perp e^{i(\bar{\kappa}_\perp - \bar{\kappa}_\perp')\bar{r}_\perp} = (2\pi)^2 \delta(\kappa_\perp - \kappa_\perp'), \tag{22}$$

$$\int \exp\left(i\frac{\omega}{c}\sqrt{\varepsilon_{1r} - \frac{\kappa_\perp^2 c^2}{\omega^2}} - i\frac{\omega'}{c}\sqrt{\varepsilon_{1r} - \frac{\kappa_\perp^2 c^2}{\omega'^2}}\right) z \, dz = \frac{2\pi c}{\varepsilon_{1r}} \sqrt{\varepsilon_{1r} - \frac{\kappa_\perp^2 c^2}{\omega^2}} \delta(\omega - \omega'). \tag{23}$$

Furthermore, note that $|E_{1\perp}^R|^2 = \frac{\omega^2}{\kappa_\perp^2 c^2}\left(\varepsilon_{1r} - \frac{\kappa_\perp^2 c^2}{\omega^2}\right)|E_{1z}^R|^2$, the expression for radiation energy is obtained as

$$W_1 = 2\int_0^{+\infty}\int \frac{q^2 \varepsilon_{1r}}{(2\pi)^3 \varepsilon_0 c \kappa_\perp^2} \sqrt{\varepsilon_{1r} - \frac{\kappa_\perp^2 c^2}{\omega^2}} |a_1|^2 d\bar{\kappa}_\perp d\omega. \tag{24}$$

We recall that $\kappa_\perp = k_1 \sin\theta = \frac{\omega}{c}\sqrt{\varepsilon_{1r}}\sin\theta$, then the integration of $d\bar{\kappa}_\perp$ is expanded as

$$d\bar{\kappa}_\perp = d\kappa_x d\kappa_y = 2\pi \kappa_\perp d\kappa_\perp = 2\pi \frac{\omega^2}{c^2}\varepsilon_{1r}\sin\theta\cos\theta \, d\theta. \tag{25}$$

The final expression for the total backward radiation energy is given as

$$W_1 = \int_0^{+\infty} W_1(\omega)\, d\omega. \tag{26-1}$$

The energy spectral density $W_1(\omega)$ of backward radiation is denoted as

$$W_1(\omega) = \int_0^{\pi/2} (2\pi \sin\theta) U_1(\omega,\theta)\, d\theta \tag{26-2}$$

The angular spectral energy density $U_1(\omega,\theta)$ of backward radiation is written as

$$\begin{aligned}
U_1(\omega,\theta) &= \frac{\varepsilon_{1r}^{3/2} q^2 \cos^2\theta}{4\pi^3 \varepsilon_0 c \sin^2\theta} |a_1|^2 \\
&= \frac{\varepsilon_{1r}^{3/2} q^2 \beta^2 \cos^2\theta \sin^2\theta}{4\pi^3 \varepsilon_0 c |1-\varepsilon_{1r}\beta^2\cos^2\theta|^2} \left| \frac{(\varepsilon_{2r}-\varepsilon_{1r})\left[1-\beta^2\varepsilon_{1r}+\beta\sqrt{\varepsilon_{2r}-\varepsilon_{1r}\sin^2\theta}\right]}{\left(1+\beta\sqrt{\varepsilon_{2r}-\varepsilon_{1r}\sin^2\theta}\right)\left[\varepsilon_{1r}\sqrt{\varepsilon_{2r}-\varepsilon_{1r}\sin^2\theta}+\varepsilon_{2r}\sqrt{\varepsilon_{1r}}\cos\theta\right]} \right|^2
\end{aligned} \tag{26-3}$$

where $\beta = \frac{v}{c}$ and $c$ is the speed of light in free space.

The angular distribution for the backward free-electron radiation, such as that in Fig. 4, can be plotted by using equation (26). As another example, Fig. S1 shows the angular spectral energy density of backward free-electron radiation (at a frequency close to the plasma frequency) for the time-domain study in Fig. 2a-b.

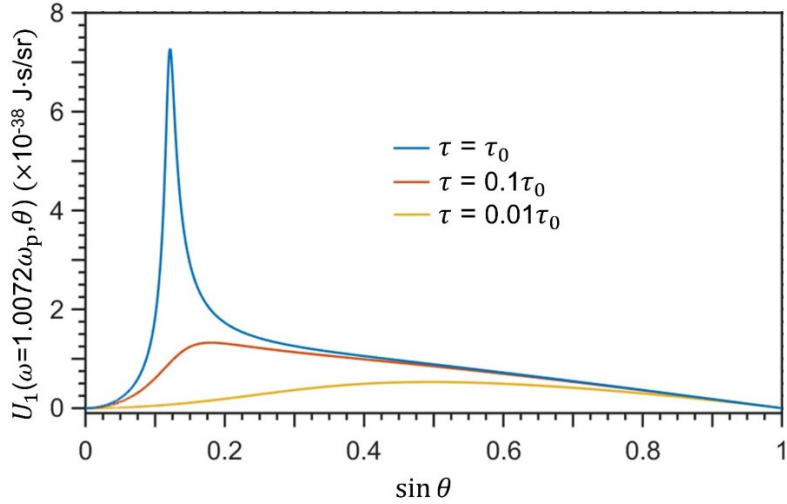

**Fig. S1. Angular spectral energy density for the time-domain study of backward free-electron radiation in Fig. 2a-b and Fig. S4 below.** Here the working frequency is set to be $\omega = 1.0072\omega_p$. Region 2 is the plasmonic medium with the relaxation time $\tau$ being $\tau_0$ (the value used in Fig. 2a-c), $0.1\tau_0$ and $0.01\tau_0$ (these two values are adopted for comparison; see also Fig. S4), respectively. When $\tau = \tau_0$, there is a peak at $\theta \approx 7°$. This peak is related to the bulk-plasmon-mediated free-electron radiation beyond the conventional formation time in Fig. 2a-b. To capture the features of bulk-plasmon-mediated free-electron radiation, we study the dynamical evolution of the backward radiation field at a point $\bar{r}_{far}$ far away from the interface but close to the electron's trajectory in Fig. 2a-b, where the angle between $\bar{r}_{far}$ and $-\bar{z}$ is $\theta_{far} \approx 7°$.

As complementary information to Fig. 4a and to see the influence of material loss, we plot in Fig. S2 the angular spectral energy density of backward free-electron radiation from the interface of the vacuum-plasmonic medium with different values of relaxation time $\tau$ for the plasmonic medium.

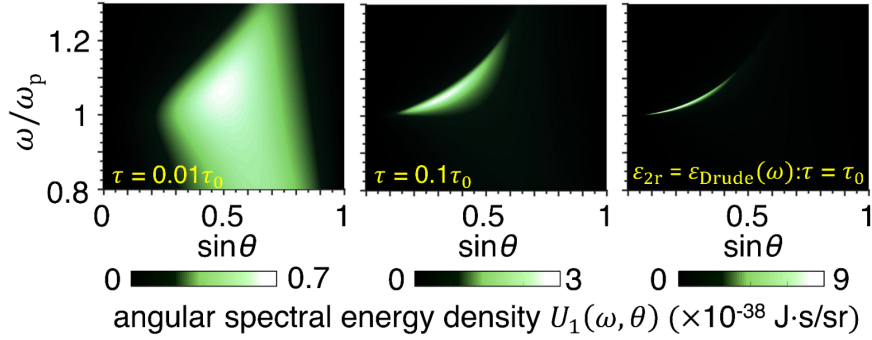

**Fig. S2. More discussion on the angular spectral energy density $U_1(\omega,\theta)$ with $\varepsilon_{2r} = \varepsilon_{Drude}(\omega)$ in Fig. 4a by using various values of $\tau$.** There is a radiation peak at each frequency (near the plasma frequency). The radiation peak has a relatively broad angular distribution if $\tau$ is small (e.g., $\theta$ ranges from 15° to 60° if $\tau = 0.01\tau_0$). If $\tau$ increases, the radiation peak becomes to have an enhanced magnitude and narrower angular distribution; see also Fig. 4a.

## *More discussion of Fig. 4a*

This sub-section serves as the complementary information for Fig. 4a. The bulk plasmons have a frequency around $\omega_p$, while the surface plasmons supported by the metal interface only exist below the frequency of $\omega_p/\sqrt{2}$. This way, the bulk plasmons and surface plasmons are independent, and they will not interact with each other. Actually, the surface plasmons are mainly formed within the conventional formation time of transition radiation. On the other hand, due to the momentum mismatch, the excited surface plasmons cannot couple into the vacuum. As a result, there is no radiation peak at the frequency (e.g. around $\omega_p/\sqrt{2}$) where surface plasmons exist, as can be seen from the backward angular spectral energy density in Fig. S3.

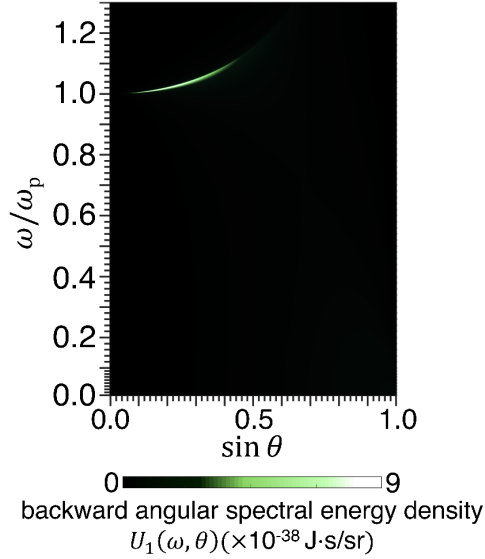

**Fig. S3. Angular spectral energy density of backward radiation.** All setups are the same as Fig. 4a. This figure serves as the complementary information for Fig. 4a. While Fig. 4a is plotted from $0.8\omega_p$ to $1.3\omega_p$, this figure is plotted in a wider frequency range, namely from 0 to $1.3\omega_p$. From this figure, there is no radiation peak below the frequency of $\omega_p/\sqrt{2}$, below which the surface plasmons are supported by the metal interface.

## Section S4. More discussion on the time-domain study of the bulk-plasmon-mediated free-electron radiation from the vacuum-plasmonic medium interface

### *More frequency-domain analysis of the bulk-plasmon-mediated free-electron radiation in Figs. 2a-c & S1*

As complementary information for Fig. 2a-c and Fig. S1 and to see the loss influence on the free-electron radiation, here we show more analysis of the free-electron radiation by setting $\tau = 0.1\tau_0$ & $0.01\tau_0$ in Fig. S4. From the time-domain and frequency-domain analyses in Fig. 2a-c, Fig. S1 and Fig. S4, after the electron crosses the interface, the radiation beyond the conventional formation time is highly dependent on the value of $\tau$ (or the material loss), especially for the frequency component near the plasma frequency. In contrast, the radiation within the conventional formation time (which can be roughly estimated as the difference between the two lines in the frequency domain in Fig. 2c and Fig. S4c & f) is slightly affected by $\tau$. This way, it is reasonable to argue that the free-electron radiation from the interface of plasmonic media within and beyond the conventional formation time is governed by different physical mechanisms. Namely, the free-electron radiation within the conventional formation time is the same as the transition radiation from regular dielectrics, and in contrast, the free-electron radiation beyond the conventional formation time is mainly due to the interaction between the long tail of bulk plasmons inside the plasmonic medium with the interface as shown in Fig. 3. We thus denote the radiation within and beyond the conventional formation time as the conventional free-electron radiation (or transition radiation) and the bulk-plasmon-mediated free-electron radiation in this work, respectively.

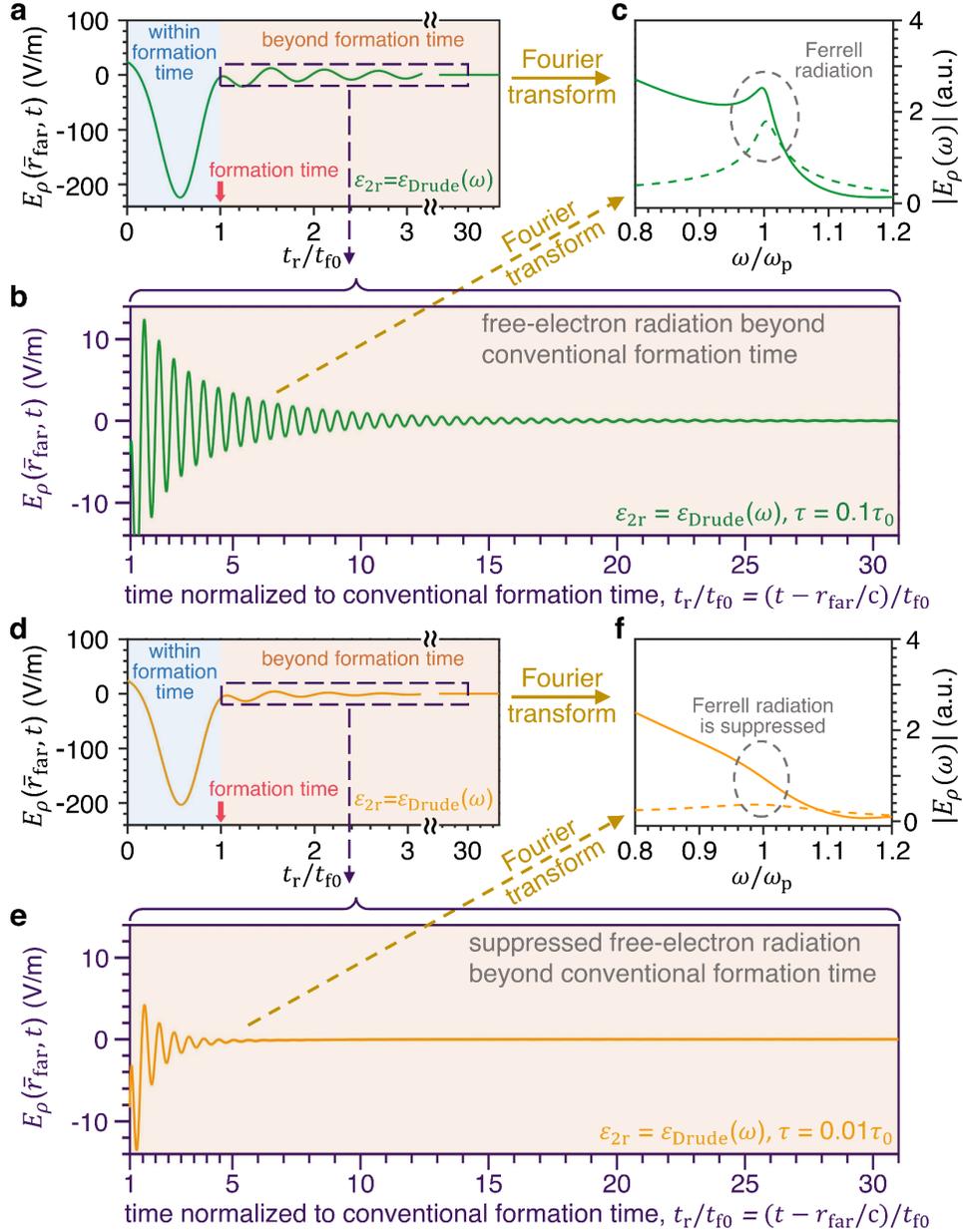

**Fig. S4. More analysis of the bulk-plasmon-mediated free-electron radiation in Fig. 2a-c by setting $\tau = 0.1\tau_0$ & $0.01\tau_0$.** All other setup is the same as Fig. 2a-c. (**a-b**) Time-domain analysis if $\tau = 0.1\tau_0$ and (**d-e**) Time-domain analysis if $\tau = 0.01\tau_0$ The bulk-plasmon-mediated free-electron radiation beyond the

conventional formation time will become more apparent if the material loss decreases or $\tau$ increases. Frequency domain analysis if (**c**) $\tau = 0.1\tau$ and (**f**) $\tau = 0.01\tau_0$. The peak of Ferrell radiation near the plasma frequency in the frequency spectrum is mainly caused by the bulk-plasmon-mediated free-electron radiation beyond the conventional formation time. If $\tau$ decreases, the peak of Ferrell radiation becomes less obvious or even vanishes (compared with that in Fig. 2c), due to the suppression of the bulk-plasmon-mediated free-electron radiation.

*More analysis on the field distribution when the electron crosses an interface in Fig. 3a-d*

As complementary information for Fig. 3a-d, the field distribution when a fast electron propagates in a homogeneous medium is shown in Fig. S5; see the related discussion in the figure caption.

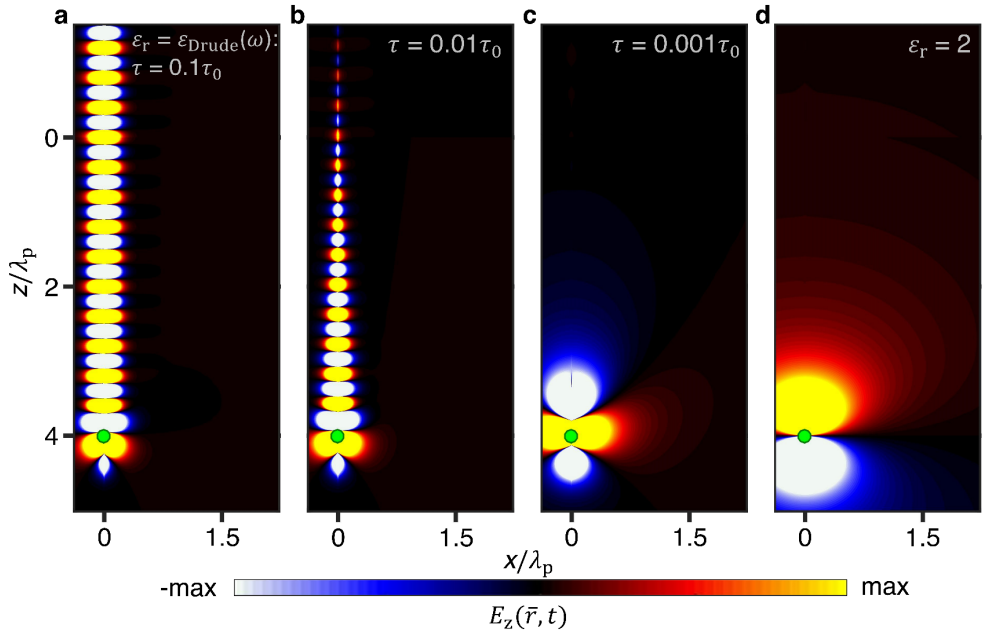

**Fig. S5. Field distribution when a fast electron moves in a homogeneous material.** (**a-c**) The homogeneous material is a plasmonic medium with $\varepsilon_r = \varepsilon_{\text{Drude}}(\omega)$. (**d**) The homogeneous material is a dielectric with $\varepsilon_r = 2$. The long tail of bulk plasmons appears only when the electron moves inside a plasmonic medium with a relatively-large relaxation time $\tau$, such as that in (a, b); the tail length increases with $\tau$. The other structural setup is the same with Fig. 3a-d.

## Section S5. History of Ferrell radiation

We introduce in this section the interesting history of Ferrell radiation since it is fraught with controversies at the very beginning. In 1958, Ferrell firstly developed an approximate theory for the radiation of plasma oscillation by pointing out that "under suitable circumstances the plasma oscillations (which can be excited by swift electrons) will give off electromagnetic radiation" [31]. Due to the importance of Ferrell radiation in the measurement of metal's plasma frequency [31], the peak in the radiation spectrum for Ferrell radiation was soon confirmed in experiments in 1960 [33,34]. However, Silin and Fetisov argued in 1961 that Ferrell radiation is just the transition radiation predicted by Ginzburg and Frank in 1945 [35]. Moreover, they considered the nonlocal response of metal and pointed out that the bulk longitudinal plasma oscillation will not cause the radiation peak. Later in 1962, Stern explained that Ferrell's method and the theory of transition radiation "are two different ways to consider the same phenomenon" [38]. Although "Ferrell's method only calculates the peak," it clearly "shows the physical mechanism causing the peak" [38]. Stern further emphasized that Silin and Fetisov misinterpreted Ferrell's physical mechanism, which is not the bulk longitudinal plasma oscillation in their study but "a surface effect" [38], i.e., "the contribution of radiative surface plasma oscillation (SPO)" [39]. (It shall be emphasized the electromagnetic fields related to the plasma oscillation mentioned in Ferrell's method [31] are transverse waves, instead of longitudinal waves.) However, Economou insisted in 1969 [39] that "there are no radiative SPO in the present geometry (i.e., a thin metal slab)" and preferred the explanation related to the transition radiation; to be specific, his explanation mainly relies on the mathematical analysis of "the denominator in the expression of transition radiation near the peak" of Ferrell radiation [39]. With the rapid development of plasmonics, it is now argued that the radiative SPO is essentially a leaky or radiative mode in the studied system and is termed as the Ferrell mode [32,38,40,44].

As emphasized in the main text, Ferrell's approximate theory and the theory of transition radiation indeed provide two seemingly distinct underlying mechanisms for Ferrell radiation, namely regarding whether it is a surface or bulk effect, from transition radiation or plasmonic oscillation, and so far there is no decisive conclusion. With our revealed mechanism of bulk-plasmon-mediated free-electron radiation beyond the conventional formation time, it becomes feasible to settle this historical debate here; see the discussion in the main text.

## Section S6. Influence of the electron's velocity on the long tail of bulk plasmons

When a swift electron moves inside a plasmonic medium with a minor material loss, the long tail of bulk plasmons will appear, as shown in Fig. 3. The bulk-plasmon tail revealed here (namely its related electromagnetic field) is transverse in nature and different from the longitudinal bulk plasmons [35,38]. To address the issue related to the longitudinal bulk plasmons, one needs to consider the nonlocal response of the plasmonic medium, which will be discussed in the next section.

Due to its transverse nature, the bulk-plasmon tail follows the dispersion relation of transverse electromagnetic waves in the plasmonic medium, i.e., $\kappa_\perp^2 + \frac{\omega^2}{v^2} = \frac{\omega^2}{c^2}\varepsilon_{\text{Drude}}$. Since $v < c/\sqrt{\varepsilon_{\text{Drude}}}$ at the frequency close to the plasma frequency, $\kappa_\perp$ will be an imaginary number (if the loss is neglected), and the bulk-plasmon tail will decay in the $\bar{r}_\perp$-direction (or in the direction perpendicular to the electron's trajectory). In other words, the bulk-plasmon tail is always well confined in space and follows the electron's trajectory inside the plasmonic medium.

Fig. S6 shows that the appearance of the bulk-plasmon tail is irrelevant to the particle velocity, but its shape will change with the particle velocity. Due to its almost periodical oscillating property in the direction of particle's motion, the tail has a wavelength in the $\bar{z}$-direction, which is around $\lambda_{z,\text{tail}} = 2\pi v/\omega_p$. Importantly, $\lambda_{z,\text{tail}}$ increases with $v$, as shown in Fig. S6.

Moreover, the bulk-plasmon-mediated free-electron radiation from the plasmonic medium is also studied for different electron's velocities in Fig. S7. Fig. S7a shows that the angular spectral energy density of backward radiation increases with the electron's velocity, but the angular position of its peak is insensitive to the electron's velocity. Figure S6b&c illustrates the time evolution of the radiated field at a point $\bar{r}_{\text{far}}$ far away from the boundary at $\theta = 7°$. We note that the appearance of the bulk-plasmon-mediated free-electron radiation beyond the conventional formation time has a minor dependence on the electron's velocity.

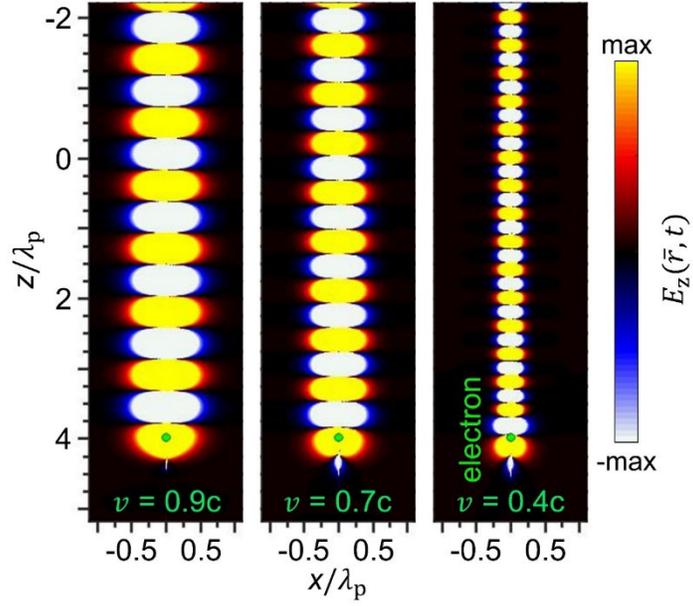

**Fig. S6. Long tail of bulk plasmons when a swift electron moves inside a homogeneous plasmonic medium with different velocities.** The relaxation time of the plasmonic medium is $\tau = 0.1\tau_0$. The bulk-plasmon tail is well confined near to the electron's trajectory, follows the electron's motion, and has a wavelength $\lambda_{z,\text{tail}}$ in the $\bar{z}$-direction, where $\lambda_{z,\text{tail}} \to 2\pi v/\omega_\text{p}$ and $\lambda_{z,\text{tail}}$ increases with $v$.

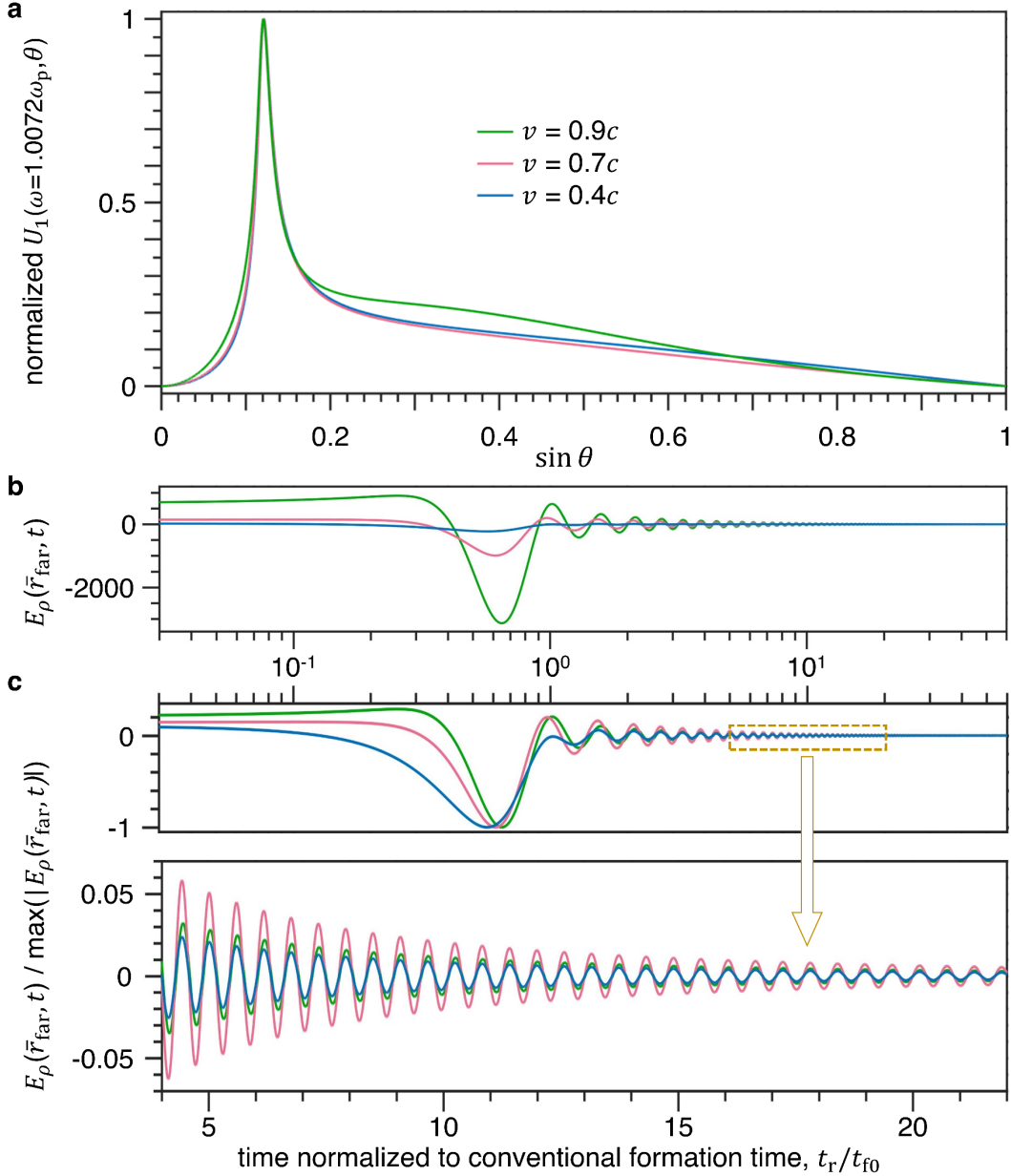

**Fig. S7. Bulk-plasmon-mediated free-electron radiation under different electron's velocities.** A swift electron, moving with different velocities, $v$, across an interface between vacuum and the plasmonic medium. The basic setup is the same as Fig. 2a-b. (**a**) Angular spectral energy density of backward radiation at a frequency close to the plasma frequency. (**b**) Dynamical evolution of radiated field at a far point $\bar{r}_{\text{far}}$ in vacuum, where the angle between $\bar{r}_{\text{far}}$ and $-\bar{z}$ is $\theta = 7°$. (**c**) Time evolution of the normalized radiation field from (b). The appearance of bulk-plasmon-mediated free-electron radiation has a minor dependence on the electron's velocity.

For the plasmonic medium, we set $\tau = \tau_0$. For the sake of clarity, the horizontal axis of all plots here are normalized by the same $t_{f0}$, namely the conventional formation time calculated by setting $v = 0.4c$.

## Section S7. Influence of the nonlocal response of plasmonic media and the longitudinal waves on the bulk-plasmon-mediated free-electron radiation

*Derivation of free-electron radiation with the consideration of nonlocal response and longitudinal waves*

When a plasmonic medium (e.g., noble metals) has its relative permittivity close to zero, the longitudinal wave might appear. As a general discussion, the longitudinal wave can be considered by including the nonlocal response or spatial dispersion of plasmonic media in our model. Below we address the influence of the nonlocal response of plasmonic media and the longitudinal waves on our revealed phenomena, by studying the angular spectral energy density of backward radiation in the frequency domain and the dynamical evolution of the radiation field in the time domain.

To consider the nonlocal response of the plasmonic medium, we adopt the hydrodynamic model [46-51]. In the hydrodynamic model, the permittivity of the medium exhibits different values for the transverse and longitudinal waves. Accordingly, the Maxwell equation is modified into

$$\left(k^2 - \frac{\omega^2}{c^2}\varepsilon^T\right)\bar{E}^T - \frac{\omega^2}{c^2}\varepsilon^L\bar{E}^L = -i\omega\mu_0(\bar{J}^T + \bar{J}^L), \tag{27}$$

where $T$ and $L$ in the superscripts denote the transverse and longitudinal components. It is worthy to note that $k^2 = \kappa_\perp^2 + k_z^2$, $\bar{k} \times (\bar{k} \times \bar{E}^T) = k^2 E^T$ for transverse waves and $\bar{k} \times \bar{E}^L = 0$ for longitudinal waves. By following a similar procedure in section S1 to solve the differential equation, the expressions for the charge field and the radiation field, for both transverse and longitudinal waves, are obtained as

$$E_z^{q(L)} = -\frac{iq\exp\left(i\frac{\omega}{v}z\right)}{\omega\varepsilon_0(2\pi)^3\left(\kappa_\perp^2 + \frac{\omega^2}{v^2}\right)}\frac{\left(\frac{\omega^2}{v^2}\right)}{\varepsilon^L}, \tag{28}$$

$$E_z^{q(T)} = \frac{iq \exp\left(i\frac{\omega}{v}z\right)}{\omega\varepsilon_0(2\pi)^3\left(\kappa_\perp^2 + \frac{\omega^2}{v^2}\right)\left[\left(\kappa_\perp^2 + \frac{\omega^2}{v^2} - \frac{\omega^2}{c^2}\varepsilon^T\right)\right]} \left[\frac{\kappa_\perp^2 \frac{\omega^2}{c^2}}{}\right], \tag{29}$$

$$E_z^{R(L)} = \frac{iq}{\omega\varepsilon_0(2\pi)^3} a^L \exp(ik_z^L z), \tag{30}$$

$$E_z^{R(T)} = \frac{iq}{\omega\varepsilon_0(2\pi)^3} a^T \exp(ik_z^T z). \tag{31}$$

The transverse wave satisfies the dispersion relation of $k^2 = \frac{\omega^2}{c^2}\varepsilon^T$, and thus $k_z^T = \sqrt{\frac{\omega^2}{c^2}\varepsilon^T - \kappa_\perp^2}$. On the other hand, the longitudinal wave satisfies the dispersion relation that $\varepsilon^L = 0$ [50], where

$$\varepsilon^L(\omega, k) = 1 - \frac{\omega_p^2}{\omega^2 + i\omega/\tau - \beta_L^2 k^2}, \tag{32}$$

In equation (32), $\beta_L = \sqrt{3/5}\, v_F$ is the hydrodynamic wave vector, and $k$ is the wave vector of the waves. This way, $k_z^L = \sqrt{k_{L0}^2 - \kappa_\perp^2}$ is related to the longitudinal permittivity, where $k_{L0}$ is the solution of $\varepsilon^L = 0$.

Next, the radiation fields in different regions, both for the transverse and longitudinal waves, can be solved by enforcing the boundary conditions. An additional boundary condition (ABC) is required to solve all unknowns. In short, the boundary conditions can be expressed as follows,

$$\left(E_{1z}^q + E_{1z}^R\right)|_{z=0} = \left(E_{2z}^{q(L)} + E_{2z}^{R(L)} + E_{2z}^{q(T)} + E_{2z}^{R(T)}\right)|_{z=0}, \tag{33}$$

$$\left(H_{1\phi}^q + H_{1\phi}^R\right)|_{z=0} = \left(H_{2\phi}^{q(T)} + H_{2\phi}^{R(T)}\right)|_{z=0}, \tag{34}$$

$$\left(\kappa_\perp E_{1\perp}^q + \kappa_\perp E_{1\perp}^R\right)|_{z=0} = \left(\kappa_\perp E_{2\perp}^{q(L)} + \kappa_\perp E_{2\perp}^{R(L)} + \kappa_\perp E_{2\perp}^{q(T)} + \kappa_\perp E_{2\perp}^{R(T)}\right)|_{z=0}. \tag{35}$$

Since region 1 is composed of a regular dielectric, all fields in region 1 are transverse. After cumbersome calculations, the amplitude for the backward radiation in region 1 is obtained as

$$a_{21}^- = \frac{1}{\left[\frac{\omega}{c}\left(\varepsilon_{1r}\sqrt{\varepsilon_{2r}^T - \frac{\kappa_\perp^2 c^2}{\omega^2}} + \varepsilon_{2r}^T\sqrt{\varepsilon_{1r} - \frac{\kappa_\perp^2 c^2}{\omega^2}}\right) + \frac{\kappa_\perp^2}{k_{z2}^L}(\varepsilon_{1r} - \varepsilon_{2r}^T)\right]} \left\{\frac{\kappa_\perp^2 \varepsilon_{2r}^T \frac{\omega}{v}\left(\varepsilon_{2r}^{L(q)} - \varepsilon_{2r}^T\right)}{\varepsilon_{2r}^{L(q)}\left(\kappa_\perp^2 + \frac{\omega^2}{v^2}\right)\left(\varepsilon_{2r}^T - \frac{c^2}{v^2} - \frac{\kappa_\perp^2 c^2}{\omega^2}\right)}\right.$$

$$+ \frac{\frac{\kappa_\perp^2 c^2}{\omega^2}\left(\varepsilon_{2r}^T \frac{\omega}{v} - \varepsilon_{2r}^{L(q)} \frac{\omega}{c}\sqrt{\varepsilon_{2r}^T - \frac{\kappa_\perp^2 c^2}{\omega^2}}\right)}{\varepsilon_{2r}^{L(q)}\left(\varepsilon_{2r}^T - \frac{c^2}{v^2} - \frac{\kappa_\perp^2 c^2}{\omega^2}\right)} - \frac{\frac{\kappa_\perp^2 c^2}{\omega^2}\left(\varepsilon_{2r}^T \frac{\omega}{v} - \varepsilon_{1r} \frac{\omega}{c}\sqrt{\varepsilon_{2r}^T - \frac{\kappa_\perp^2 c^2}{\omega^2}}\right)}{\varepsilon_{1r}\left(\varepsilon_{1r} - \frac{c^2}{v^2} - \frac{\kappa_\perp^2 c^2}{\omega^2}\right)}$$

$$+ \frac{\kappa_\perp^2}{k_{z2}^L}\left[\frac{\varepsilon_{2r}^T\left(\varepsilon_{2r}^{L(q)}\kappa_\perp^2 + \varepsilon_{2r}^T \frac{\omega^2}{v^2}\right)}{\varepsilon_{2r}^{L(q)}\left(\kappa_\perp^2 + \frac{\omega^2}{v^2}\right)\left(\varepsilon_{2r}^T - \frac{c^2}{v^2} - \frac{\kappa_\perp^2 c^2}{\omega^2}\right)} - \frac{\varepsilon_{2r}^{L(q)} \frac{\kappa_\perp^2 c^2}{\omega^2} + \varepsilon_{2r}^T \frac{c^2}{v^2}}{\varepsilon_{2r}^{L(q)}\left(\varepsilon_{2r}^T - \frac{c^2}{v^2} - \frac{\kappa_\perp^2 c^2}{\omega^2}\right)} + \frac{\varepsilon_{1r}\frac{\kappa_\perp^2 c^2}{\omega^2} - \varepsilon_{2r}^T\left(\varepsilon_{1r} - \frac{c^2}{v^2}\right)}{\varepsilon_{1r}\left(\varepsilon_{1r} - \frac{c^2}{v^2} - \frac{\kappa_\perp^2 c^2}{\omega^2}\right)}\right]\right\}.$$
(36)

One should bear in mind that $\varepsilon_{2r}^{L(q)}(\omega, k) = 1 - \frac{\omega_p^2}{\omega^2 + i\omega/\tau - \beta_L^2(\omega^2/v^2 + \kappa_\perp^2)}$ in equation (36), where $k^2 = \kappa_\perp^2 + \omega^2/v^2$ and $\omega/v$ is the $\bar{z}$-component of wavevector for the charge field.

### *Influence of the nonlocal response and longitudinal waves on the field distribution of free-electron radiation*

Figure S7 studies the influence of the nonlocal response of the plasmonic medium and the excitation of longitudinal waves on the field distribution of free-electron radiation. We find the nonlocal response have a minor influence on the long tail of bulk plasmons inside the plasmonic medium and the radiation fields inside the region of free space; see Fig. S8a-c. Therefore, it is reasonable to argue that the nonlocal response have a negligible influence on the bulk-plasmon-mediated free-electron radiation revealed in this work; see more analysis in Fig. S9. Moreover, in addition to the long tail of transverse bulk plasmons, if we consider the nonlocal response, the Cherenkov radiation of longitudinal waves will also be excited when the electron moves inside the plasmonic medium; see Fig. S8d-e. Since the excited longitudinal waves inside the plasmonic medium propagate forward (i.e., away from the interface) and cannot propagate far, their appearance also has a minor influence on the bulk-plasmon-mediated free-electron radiation revealed in this work.

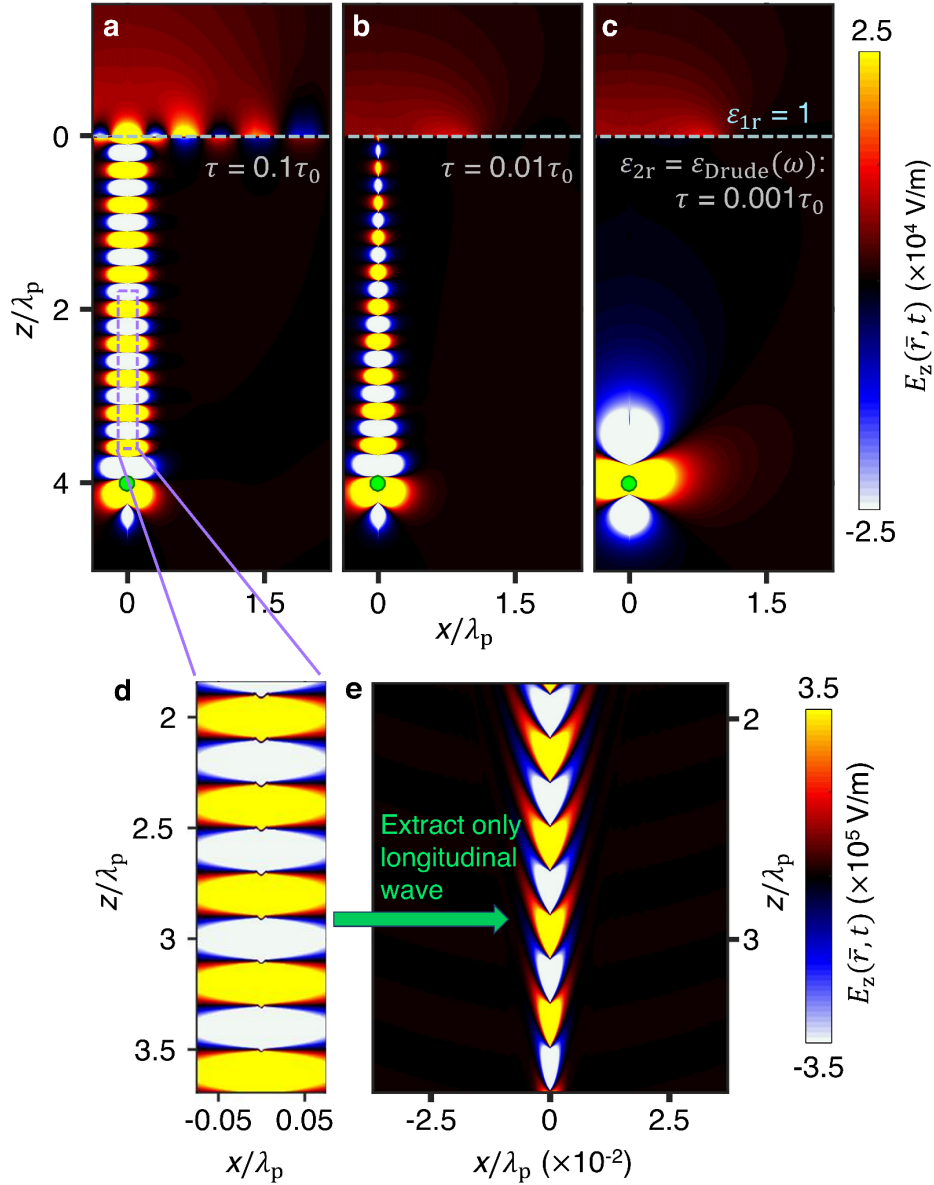

**Fig. S8. Influence of the nonlocal response and longitudinal waves on the distribution of total field when the electron perpendicularly crosses an interface of the vacuum-plasmonic medium. (a-c)** Distribution of the total field with the same structural setup as Figs. 3a-c and S4. **(d)** The magnified view of a small region is highlighted by the dashed square in (a). The plots in (a-d) contain both longitudinal and transverse waves. **(e)** Extraction of the longitudinal wave from (d); in other words, the plot in (e) only contains the longitudinal wave. While Figs. 3a-c and S4 have neglected the nonlocal response of plasmonic media, here we study its influence. The nonlocal response of $\beta_L$ is assumed to be $c/300$ (the value generally obtained from silver). Compared with

Fig. 3a-c, Fig. S8a-c shows that the nonlocal response of plasmonic media has a minor influence on the radiation fields in the region of free space and the long tail of bulk plasmons inside the plasmonic medium. These figures indicate that the nonlocal response has little influence on the radiation spectrum of backward radiation and the revealed bulk-plasmon-mediated free-electron radiation; see more in Fig. S9. Besides, if we consider the nonlocal response, there will be some eigenmodes of longitudinal waves inside the plasmonic medium. This way, in addition to the long tail of transverse bulk plasmons mentioned in Fig. 3a and S4a, there will also be the emergence of Cherenkov radiation of plane-like longitudinal waves inside the plasmonic medium. The magnitude of longitudinal waves is relatively weak and they (see the little spikes in the long tail of bulk plasmons in Fig. S8d) are flooded by the long tail of transverse bulk plasmons; as a result, the longitudinal waves cannot be easily seen in the total field plots in Fig. S8a-c. To visualize the excited longitudinal waves clearly, we plot them solely in Fig. S8e. The excited longitudinal waves propagate away from the interface. Due to the material loss, they cannot propagate over a long distance or far away from the electron trajectory.

***Influence of the nonlocal response and longitudinal waves on the angular spectral energy density and the bulk-plasmon-mediated free-electron radiation***

Fig. S9a demonstrates the influence of the nonlocal response of plasmonic media and the excitation of longitudinal waves on the angular spectral energy density of backward radiation from the interface of the vacuum-plasmonic medium. Although the nonlocal response slightly affects the magnitude of the peak, the peak is still pronounced in Fig. S9a. Furthermore, the time-domain study of bulk-plasmon-mediated free-electrons radiation in Fig. S9b confirms the minor influence of the nonlocal response on the bulk-plasmon-mediated free-electron radiation.

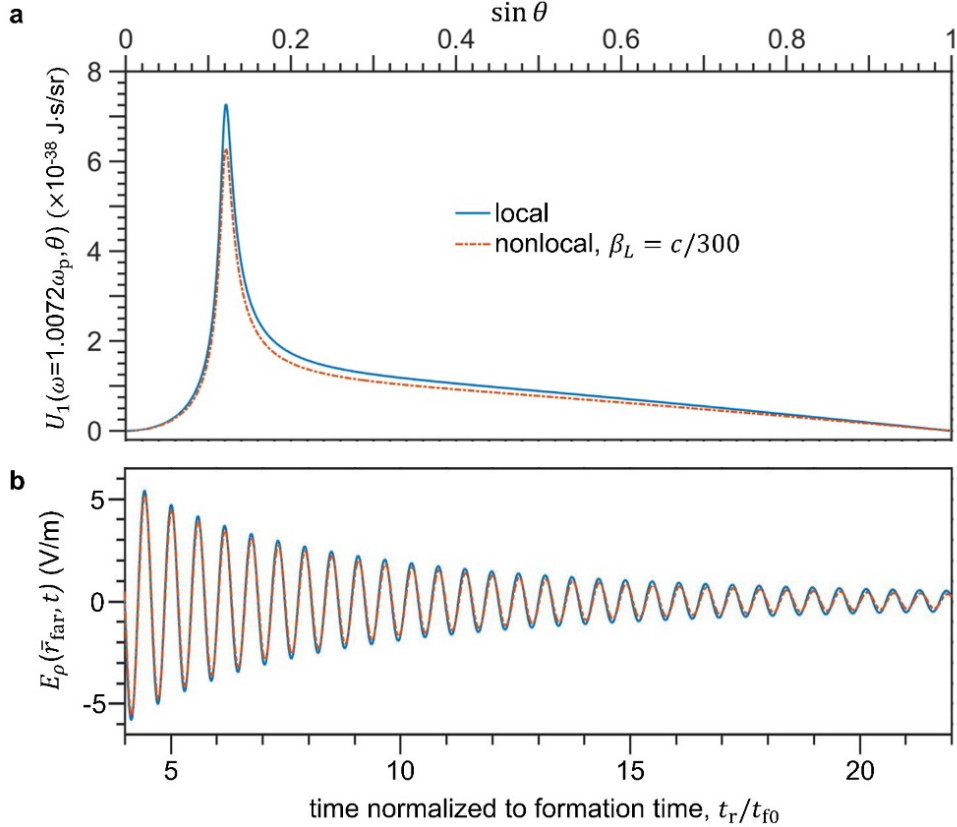

**Fig. S9. Influence of nonlocal response on the bulk-plasmon-mediated free-electron radiation.** A swift electron passes an interface between free space and a plasmonic medium. The basic setup is the same as Fig. 2 and the plasmonic medium has $\tau = \tau_0$. (**a**) Angular spectral energy density of backward radiation at a frequency close to the plasma frequency. (**b**) Time evolution of the radiated field at a point far from the interface, where the angle between $\bar{r}_{\text{far}}$ and $-\bar{z}$ is $\theta = 7°$. The hydrodynamic wave vector $\beta_L$ characterizes the nonlocal response, where the nonlocal response increases with $\beta_L$. As a typical example, $\beta_L = c/300$ (which is the one used for the study of the nonlocal response of silver) is adopted. The nonlocal response has a minor influence on the bulk-plasmon-mediated free-electron radiation.

## Supplementary References

## Caption for Supplementary Video

Movie S1. Dynamical evolution of $E_z(\bar{r}, t)$ field distribution in Fig. 3a.

Movie S2. Dynamical evolution of $E_z(\bar{r}, t)$ field distribution in Fig. 3b.

Movie S3. Dynamical evolution of $E_z(\bar{r}, t)$ field distribution in Fig. 3c.

Movie S4. Dynamical evolution of $E_z(\bar{r}, t)$ field distribution in Fig. 3d.

Movie S5. Field distribution when a charge field moves inside a plasmonic medium with various values of relaxation time $\tau$. The other structural setup is the same as Fig. S5.